\catcode`\@=11
 
\def\citen#1{\if@filesw \immediate\write \@auxout {\string\citation{#1}}\fi%
\@tempcntb\m@ne \let\@h@ld\relax \def\@citea{}%
\@for \@citeb:=#1\do {\@ifundefined {b@\@citeb}%
    {\@h@ld\@citea\@tempcntb\m@ne{\bf ?}%
    \@warning {Citation `\@citeb ' on page \thepage \space undefined}}%
    {\@tempcnta\@tempcntb \advance\@tempcnta\@ne
    \setbox\z@\hbox\bgroup\ifcat0\csname b@\@citeb \endcsname \relax
    \egroup \@tempcntb\number\csname b@\@citeb \endcsname \relax
    \else \egroup \@tempcntb\m@ne \fi \ifnum\@tempcnta=\@tempcntb
    \ifx\@h@ld\relax \edef \@h@ld{\@citea\csname b@\@citeb\endcsname}%
    \else \edef\@h@ld{\hbox{--}\penalty\@highpenalty
    \csname b@\@citeb\endcsname}\fi
    \else \@h@ld\@citea\csname b@\@citeb \endcsname \let\@h@ld\relax \fi}%
\def\@citea{,\penalty\@highpenalty\hskip.13em plus.13em minus.13em}}\@h@ld}
\def\@citex[#1]#2{\@cite{\citen{#2}}{#1}}%
\def\@cite#1#2{\leavevmode\unskip\ifnum\lastpenalty=\z@\penalty\@highpenalty\fi%
  \ [{\multiply\@highpenalty 3 #1%
  \if@tempswa,\penalty\@highpenalty\ #2\fi}]}   %
\makeatother 

\def\alg           {algebra}
\def\be            {\begin{equation}}
\def\bearl         {\begin{array}{l}}
\def\bearll        {\begin{array}{ll}}
\def\bearlll       {\begin{array}{lll}}
\def\bfe           {{\bf1}}
\def\cala          {{\liefont A}}
\def\calap         {\tilde{\liefont A}}
\def\calg          {{\cal G}}
\def\caln          {{\cal N}}
\def\calr          {{\cal R}}
\def\calrp         {\tilde{\cal R}}
\def\calu          {{\cal U}}
\def\calno         {\chec{\cal N}}
\def\calnp         {\tilde{\cal N}}
\def\cb            {{\cal B}}
\def\cft           {conformal field theory}
\def\cfts          {conformal field theories}
\def\chec          {\breve }
\def\chii          {\raisebox{.15em}{$\chi$}}
\def\complex       {{\dl C}}

\def\cza           {charge-zero algebra} 

\def\dl            {\mathbb }

\def\ee            {\end{equation}}
\def\eE            {{\rm e}}
\def\eear          {\end{array}}
\def\eq            {\,{=}\,}
\newcommand\erf[1] {(\ref{#1})}
\def\findim        {finite-dimensional}
\newcommand\fline[1]{\vfill\noindent ------------------\\[1 mm]}
\newcommand\Frac[2]{\mbox{\large$\frac{#1}{#2}$}}
\def\frc           {fusion rule coefficient}
\def\fsa           {classifying algebra}
\def\futnote#1     {\footnote{~#1}\ }

\def\g             {\mbox{$\liefont g$}}
\def\gM            {{\liefont g}}

\def\hw            {highest weight}

\def\ii            {{\rm i}}
\def\iN            {\,{\in}\,}
\def\irrep         {irreducible representation}
\def\J             {{\rm J}}
\long\def\labl#1   {\label{#1}\ee}
\def\lie           {Lie algebra}
\def\liefont       {\mathfrak }

\def\llb           {\mbox{\large(}}
\def\lrb           {\mbox{\large)}}
\def\mi            {\,{-}\,}
\def\Nf            {N_{\rm f}}

\def\P             {^+_{\phantom I}}

\def\pl            {\,{+}\,}
\def\rep           {representation}
\long\def\query#1{\hskip 0pt{\vadjust{\everypar={}\small\vtop to 0pt{\hbox{}%
     \vskip -13pt\rlap{\hbox to 49.0pc{\hfil{\vtop{\hsize=8pc\tolerance=6000%
     \hfuzz=.5pc\rightskip=0pt plus 3em\noindent#1}}}}\vss}}}}%
\newcommand\sect[1]{\section{#1}} 

\def\su            {$\liefont{su}$(2)}
\newcommand\sumfix[1]{\sum_{#1:\;\J #1=#1}}
\newcommand\sumhalf[1]{\sum_{#1:\;Q(#1)=1/2}}
\newcommand\sumnull[1]{\sum_{#1:\;Q(#1)=0}}
\def\tj            {\Theta_\J}
\newcommand\version[1] {\ifnum\draftcontrol=1 \typeout{}\typeout{#1}\typeout{}
                   \vskip3mm \centerline{\fbox{{\tt DRAFT -- #1 -- }
                   {\small\draftdate}}}
                   \vskip3mm \fi}
\def\wzwm          {WZW model}
\def\wzwt          {WZW theory}
\def\wzwts         {WZW theories}
\def\zet           {{\dl Z}}
\def\zetpluso      {\mbox{$\zet_{\ge0}$}}

\def\draftdate{\number\month/\number\day/\number\year\ \ \ \hourmin }

\global\def\draftcontrol{0}
\catcode`\@=12
 
\documentclass[12pt]{article} \usepackage{amssymb,amsfonts}
\begin{document}

 
\begin{flushright}  {~} \\[-15 mm]  {\sf hep-th/9708141} \\[1mm]
{\sf CERN-TH/97-215} \\[1 mm] 
{\sf August 1997} \end{flushright}
 
\begin{center} \vskip 13mm
{\Large\bf A CLASSIFYING ALGEBRA}\\[4mm]
{\Large\bf FOR BOUNDARY CONDITIONS}\\[16mm]
{\large J\"urgen Fuchs} $^{\sf X}$ \\[3mm]
DESY\\[.6mm] Notkestra\ss e 85, \ D -- 22603~~Hamburg\\[11mm]
{\large Christoph Schweigert} \\[3mm] CERN \\[.6mm] CH -- 1211~~Gen\`eve 23
\end{center}
\vskip 20mm
 
\begin{quote}{\bf Abstract}\\[1mm]
We introduce a finite-dimensional algebra that controls the possible boundary 
conditions of a conformal field theory. For theories that are obtained by 
modding out a $\zet_2$ symmetry (corresponding to a so-called 
$D_{\rm odd}$-type, or half-integer spin simple current, modular invariant),
this classifying algebra contains the fusion algebra of the untwisted sector
as a subalgebra. Proper treatment
of fields in the twisted sector, so-called fixed points, leads to structures 
that are intriguingly close to the ones implied by modular invariance for 
conformal field theories on closed orientable surfaces.
\end{quote}
\vfill {}
\begin{flushleft}  {~} \\[-3 mm]  
{\sf CERN-TH/97-215} \\[1 mm] 
{\sf August 1997} \end{flushleft}
\fline{} {\small $^{\sf X}$~~Heisenberg Fellow}

\sect{Introduction}

Recently, theories of open strings and conformal field theories on surfaces with
boundaries have received considerable interest. As exemplified by the r\^ole 
played by D-branes in the description of non-perturbative aspects of string 
theory, it is a crucial task to obtain more 
insight into the possible boundary conditions for such theories. So far, 
however, most investigations have been limited to models based on free field 
theories or on orbifolds of such theories. In this paper we investigate the 
structure of boundary conditions in a general \cft, including non-trivial 
modular invariants.

A \cft\ typically admits several modular invariants. One always
has the charge conjugation and the diagonal modular invariant. The possible
boundary conditions in a theory with charge conjugation modular invariant
have been explored in \cite{card9}. A first investigation of non-trivial
modular invariants has been undertaken in \cite{prss2,prss3} 
for \wzwm s based on \su. The goal of this letter is to extend this work to
arbitrary rational \cfts\ with a specific type of non-diagonal modular 
invariant.

The type of modular invariant we will focus on generalizes the 
modular invariant of $D_{\rm odd}$-type in the $A$-$D$-$E$ classification of 
\su\ modular invariants, see equations \erf{inv} and \erf{23}
below. Quite generally, every non-trivial modular invariant can be 
obtained as follows: one first extends the chiral algebra and then superposes
an automorphism of the fusion rules. The invariants \erf{inv} and \erf{23} 
provide examples of non-trivial fusion rule automorphisms.
The extension procedure is by now fairly
well understood, at least in the case of extensions by so-called simple 
currents, and  can be described entirely in terms of a chiral half of the 
theory. As a consequence, such extensions do not raise any problems in the 
construction of open string theories that were not already encountered for 
closed strings. In contrast, the implementation of fusion rule automorphisms is
far from being understood. Our results provide new insight into this problem.

The further layout of this letter is as follows. In section 3 we discuss
the reflection coefficients which characterize the various consistent boundary 
conditions and conclude that the construction of an open string theory 
requires a certain relation between the numbers of primary fields of various 
types. This non-trivial relation, given in \erf{srule} below, is established 
in section 4. These results enable us to introduce in section 5 a new \findim\ 
algebra, which in section 6 is shown to control the boundary conditions. In 
the same section we also study the implications for the annulus amplitude. In 
the last section we point out possible further consequences of our results.

\sect{The modular invariant}

We analyze \cfts\ that are characterized by modular invariant
combinations of characters that generalize the non-diagonal modular invariants
  \be  Z = \sum_{l=0}^{k/2} |\chii_{2l}^{}|^2 +
  \sum_{l=0}^{k/2-1} \chii_{2l+1}^{}\chii_{k-2l-1}^*   \labl{inv}
of the \su\ \wzwt, which exist at all levels $k$ with $k=2\bmod 4$.
Here we have labelled the primary fields by their \hw\ $\Lambda$, i.e.\ by 
twice their isospin. Notice that the primary fields come in two groups: 
when the \su\ \rep s carried by a field have integral isospin, then the
\rep\ is paired with itself, while a representation with half-integral isospin 
$l$ gets paired with the representation of isospin $k/2-l$. Note that
the transition from $l$ to $k/2-l$ corresponds to taking the
fusion product with the primary field of \hw\ $k$, which is a so-called 
{\em simple current\/}: $\phi_k^{}\ast\phi_\Lambda^{}=\phi_{k-\Lambda}^{}$.

One can think of the modular invariant \erf{inv} as being obtained from the 
diagonal modular invariant by modding out a $\zet_2$ symmetry; the first sum 
in \erf{inv} then constitutes the partition function of the untwisted sector,
while the second sum is the contribution from the twisted sector. Note that 
the partition function of the twisted sector 
contains one term which superficially looks like an untwisted term, namely 
$\chii_{k/2}^{}\chii_{k/2}^*$. The corresponding primary field
has the property that it equals its fusion product with the simple current,
  $\phi_k^{}\ast\phi_{k/2}^{}=\phi_{k/2}^{}$;
it is therefore termed a {\em fixed point}. In short, the primary fields in 
our example can be organized in {\em three\/} different types: we have 
$N_0\eq k/2{+}1$ left-right symmetric integer isospin fields, $N_1\eq k/2{-}1$  
left-right asymmetric half-integer isospin fields that are not fixed points,
and $\Nf\eq1$ fixed point.

The situation summarized above has the following generalization. We consider a
rational \cft\ which contains a simple current, i.e.\ a primary field \J\ 
whose fusion product with every primary field\,%
\futnote{The primary fields are now characterized by suitable labels
$\Lambda$; in the special case of \wzwts\ these correspond to integrable 
\hw s of the underlying affine \lie. Also, we reserve the label $0$ to stand 
for the identity (vacuum) primary field, $\phi_0\equiv\bfe$.}
$\phi_\Lambda^{}$ contains just a single 
primary field, which we denote by $\phi_{\J\Lambda}^{}=\J\ast\phi_\Lambda^{}$. 
Furthermore, we assume that the simple current \J\ has order $2$, i.e.\
satisfies $\J\ast\J=\phi_0$, and that its conformal weight is half-integral, 
$\Delta_\J\iN\zet{+}1/2$.
To each primary field $\phi_\Lambda$ one associates its {\em monodromy charge\/}
$Q(\Lambda)$ with respect to \J, which is the combination
  \be Q(\Lambda) := \Delta_\J + \Delta_\Lambda - \Delta_{\J\Lambda} 
  \ \bmod\zet \labl{qdef}
of conformal weights; the monodromy charge $Q$ generalizes the conjugacy class 
-- integral or half-integral isospin -- of the \su\ example \erf{inv}. An 
important property of $Q$ is that it is conserved under operator products. 
It also appears in the relation \cite{scya6}
  \be  S_{\J\lambda,\mu} = \eE^{2\pi\ii Q(\mu)}\, S_{\lambda,\mu}  \labl{sm}
for the modular transformation matrix $S$.

It is known \cite{scya6} that in the situation at hand the following 
non-diagonal combination of characters is modular invariant:
  \be Z= \sumnull\Lambda \chii_\Lambda^{}\chii_{\Lambda^+}^* +
  \sumhalf\Lambda \chii_\Lambda^{} \chi_{(\J\Lambda)\P}^* \, . \labl{23}
Here $\lambda^+$ denotes the label of the field $\phi_{\lambda\P}^{} \equiv( 
\phi_{\lambda}^{})^+_{}$ that is the charge conjugate of $\phi^{}_{\lambda}$.\,%
\futnote{Let us remark that
usually instead of \erf{23} one considers the combination of characters
where the charge conjugation is absent, which is modular invariant as well.
(This type of modular invariant arises naturally in the \cft\ description of 
type IIA compactifications of the superstring.) As will become clear below, in 
the open string context it is more natural to include the charge conjugation as
in \erf{23}.}

It follows from \erf{qdef} that the primary fields $\phi_\lambda$ which obey
$\J\ast\phi_\lambda=\phi_\lambda$, i.e.\ the fixed points, all have monodromy 
charge $Q(\lambda)=\Delta_\J \bmod\zet =1/2$. Also, we can again organize the 
primary fields in three different sets, the $Q=0$ fields, the $Q=1/2$ fields 
that are not fixed points, and the fixed points, with $N_0$, $N_1$ and $\Nf$ 
elements, respectively. Finally, we can again regard the invariant \erf{23} as 
being obtained from the charge conjugation invariant by modding out 
the $\zet_2$ symmetry that is induced by the simple current \J.
In this picture, the fixed points, even though left-right symmetric, 
all belong to the twisted sector; in the investigations below, this always 
must be kept in mind.

\sect{Boundary conditions and reflection coefficients}

Let us now analyze such \cfts\ on surfaces with boundaries. Throughout this 
letter we assume that the boundary conditions are not only compatible with 
conformal invariance, but that they even preserve the full 
chiral symmetry of the theory. This condition effectively links left and right
movers, and as a consequence a primary field in the bulk can survive 
in the presence of a boundary only if in the torus partition function
it is paired with its charge conjugate. In the case of the modular invariant 
\erf{23} this condition is fulfilled for the $N_0$ fields with vanishing 
monodromy charge and for the $\Nf$ fixed points, so that we are left with 
$N_0+\Nf$ bulk fields.

The investigation of \cfts\ in the presence of boundaries is based on the fact 
that every surface with boundaries admits a twofold cover that is orientable 
and does not have any boundaries. Under the lift to the covering surface
points in the bulk have two pre-images, while for boundary points the lift is 
unique. It follows that when a bulk field $\phi_{(\lambda,\lambda\P)}$
approaches a boundary, one effectively
has to take the operator product of fields sitting at the two pre-images on 
the covering surface, and as a consequence it excites boundary fields
$\psi^{\alpha\alpha}_\mu$ \cite{lewe3}. On the (unit) disk, this is encoded 
in the expansion 
  \be  \phi_{(\lambda,\lambda\P)}(r\eE^{\ii\sigma})
  \,\sim\, \sum_{\mu,\alpha} C^\alpha_{(\lambda,\lambda\P),\mu}\, 
  (1\mi r^2)^{-2\Delta_\lambda+\Delta_\mu}_{}\,
  \psi^{\alpha\alpha}_\mu(\eE^{\ii\sigma}) \qquad {\rm for}\;\ r\to 1 \,.  \ee
Here the possible boundary conditions are labelled by $\alpha$. 

The constants $C^\alpha_{(\lambda,\lambda\P),\mu}$ can be interpreted as
reflection coefficients at a boundary $\alpha$ with excitation of type $\mu$;
the determination of their explicit values is one of the necessary ingredients
for formulating a \cft\ on surfaces with boundaries. For all theories studied 
so far it was found that the reflection coefficients 
$C^\alpha_{(\lambda,\lambda\P),0}$ involving the identity boundary
field form \irrep s of some semisimple \findim\ algebra $\calap$. Accordingly 
we call $\calap$ the {\em \fsa\/}.
In the case of the charge conjugation modular invariant it was argued in 
\cite{card9} that this algebra $\calap$ just coincides with the fusion rule 
algebra $\cala$ of the theory, whose structure constants are the \frc s 
$\caln_{\lambda,\mu}^{\ \ \nu}$; a distinguished basis of $\cala$ is given by 
all bulk fields, so that one has the relation
  \be  C^\alpha_{(\lambda,\lambda\P),0} C^\alpha_{(\mu,\mu\P),0}=
  \sum_\nu \caln_{\lambda,\mu}^{\ \ \nu}\, C^\alpha_{(\nu,\nu\P),0} \,. \ee

A similar structure was found in \cite{prss3} for the \su\ \wzwt\ 
with the modular invariant \erf{inv}. Again the possible boundary conditions
can be related to the \irrep s of a \findim\ semisimple associative algebra 
$\calap$ which possesses a basis labelled by the allowed bulk fields; in 
particular the dimension of $\calap$ is now
$N_0+\Nf$. The structure constants of this algebra have
been determined in \cite{prss3} by using the explicit form of the duality 
(i.e., fusing and braiding) matrices of these models. (Besides the $c\,{<}\,1$ 
minimal models, the \su\ \wzwm s\ are actually the only \cfts\ for which 
explicit closed expressions for the duality matrices are known.) In this letter
we will generalize the arguments of \cite{card9} to arrive at a general 
prescription for the \fsa\ $\calap$ for {\em all\/} modular invariants of the 
type \erf{23}.

It was also observed in \cite{prss3} that the number of boundary conditions,
which equals the number of \irrep s of $\calap$, is given by
  $k/2+2=\frac12(N_0+N_1)+2\Nf$;
thus, provided one counts length-1 (i.e.,
fixed point) orbits with a multiplicity 2, the \irrep s of $\calap$ are in
one-to-one correspondence with the orbits of \J.  Analogously we
expect a total of $\frac12(N_0+N_1)+ 2 \Nf$ possible boundary 
conditions also in the general case \erf{23}. Since the number of \irrep s of 
a semisimple algebra is equal to its dimension, this can hold only if the 
numbers of primary fields of different types satisfy the non-trivial relation
  \be  \Frac12\,(N_0+N_1)+ 2 \Nf = N_0+\Nf \,.  \labl{srule}
In the \su\ WZW case this identity is obviously valid. As a first step towards 
a formula for the reflection coefficients, we now show that the sum rule 
\erf{srule} holds in general.

\sect{The charge-zero subalgebra of the fusion algebra}

We will establish the sum rule \erf{srule} by investigating the properties of 
a certain associative \alg\ $\cala_0$. We introduce $\cala_0$ as the subalgebra
of the fusion algebra $\cala$ that is spanned by the $N_0$ primary fields with 
vanishing monodromy charge $Q$. It is easily checked that $\cala_0$ inherits 
from $\cala$ the structure of a fusion algebra, so that in particular it is 
semisimple and its dimension $N_0$ equals the number of its \irrep s. We claim 
that this number is equal to the number of orbits of \J, i.e.\ that
  \be  N_0 = \Frac12\, (N_0 + N_1) + \Nf \,,  \labl{41}
which is equivalent to the sum rule \erf{srule}.

To derive \erf{41}, we first observe that the monodromy charge 
$Q$ induces a $\zet_2$ grading on the fusion \alg\ $\cala$. As a consequence, 
the representation matrices of $Q\eq0$ primary fields in the regular \rep\,%
\futnote{The \rep\ matrices $\caln_\lambda$ in this \rep\ are just the fusion 
matrices, i.e.\ $(\caln_\lambda)_\mu^{\ \nu}=\caln_{\lambda,\mu}^{\ \ \nu}$.}
of $\cala$ are of block diagonal form; more
precisely, they are built from two blocks, one of them being the representation
matrix of the field in the regular representation of $\cala_0$. It follows that
the characteristic polynomials of fusion matrices for such primary fields
factorize and contain as a factor the characteristic polynomial in the
regular \rep\ of $\cala_0$. Since the roots of these polynomials are just the 
\rep\ matrices of the (one-dimensional) \irrep s, we conclude that every
irreducible representation of the \cza\ $\cala_0$ is obtained by restricting 
an \irrep\ of $\cala$.
 
Now the \irrep s $\calr_\alpha{:}\;\cala\,{\to}\,\complex$ of $\cala$, the 
so-called generalized quantum dimensions, are in one-to-one correspondence to 
the primary fields; they are expressible through the modular matrix $S$:
  \be  \calr_\alpha(\phi_\mu) = \frac{S_{\alpha,\mu}}{S_{0,\alpha}} \,.  \ee
As a consequence of the simple current relation \erf{sm}, the restrictions
of the $\calr_\alpha$ to the sub\alg\ $\cala_0$ coincide whenever the
labels $\alpha$ belong to one and the same orbit of \J. This implies that 
the dimension of $\cala_0$ is smaller or equal to the number of orbits of \J.
 
It remains to be shown that two \irrep s of the fusion \alg\ $\cala$
take the same value on all fields of vanishing monodromy charge {\em only\/}
if they are related by the action of the simple current, since this implies 
that the dimension of $\cala_0$ is
larger or equal to the number of orbits. To this end, we use the fact that the 
modular matrix $S$ is unitary, which when combined with \erf{sm} yields 
  \be \delta_{\alpha,\beta}^{}= \sum_\mu S_{\alpha,\mu}^{} S^*_{\mu,\beta} 
  \qquad\mbox{and}\qquad
  \delta_{\J\alpha,\beta}^{}= \sum_\mu S_{\J\alpha,\mu}^{} S^*_{\mu,\beta}
  = \sum_\mu \eE^{2\pi\ii Q(\mu)} S_{\alpha,\mu}^{} S^*_{\mu,\beta} \,, \ee
from which we conclude that
  \be \sumnull\mu S_{\alpha,\mu}^{} S^*_{\mu,\beta} =
  \Frac12\,(\delta_{\alpha,\beta}^{}+\delta_{\J\alpha,\beta}^{}) \,.  \labl{45}
On the other hand, if 
two \irrep s $\calr_\alpha$ and $\calr_\beta$ of the fusion algebra 
coincide (i.e.\ if $S_{\alpha,\mu}/S_{0,\alpha}\eq S_{\beta,\mu}/S_{0,\beta}$)
for all $\mu$ with $Q(\mu)\eq0$, then we have $S_{\alpha,\mu}\eq\lambda 
S_{\beta,\mu}$ with $\lambda\neq 0$ for all those $\mu$. As a consequence,
  \be \Frac12\,(\delta_{\alpha,\beta}^{} + \delta_{\J\alpha,\beta})^{}
  = \!\sumnull\mu S_{\alpha,\mu}^{} S^*_{\mu,\beta}
  = \lambda \!\sumnull\mu S_{\beta,\mu}^{}S^*_{\mu,\beta}
   =\Frac12\,(1+\delta_{\J\beta,\beta})^{}\,\lambda \,,  \ee
which shows that $\lambda=1$ and that $\alpha$ and $\beta$ are on the same
simple current orbit. This concludes our derivation of the relation \erf{41}, 
and hence of \erf{srule}.
 
\smallskip
As a side remark, we mention the following generalization of the structure 
discovered above.  We denote by $\calu$ any subgroup of the (abelian) group 
$\calg$ of all simple currents of a \cft, and by $\cala_0$ the sub-fusion 
algebra that is spanned by all fields whose monodromy charges with respect to 
all simple currents in $\calu$ vanish. Then the number of orbits of $\calu$ 
on $\cala$ is just $N_0$, the dimension of $\cala_0$. To see this, we observe 
that by the same arguments as before the \cza\ $\cala_0$ is semisimple. 
Moreover, again one has a grading of $\cala$ (this time over the group 
$\calg/\calu$), leading to block diagonal fusion matrices 
for primary fields in the subspace $\cala_0$. The formula \erf{sm} still 
guarantees that the generalized quantum dimensions belonging to fields on one 
and the same orbit of $\calu$ give one and the same \irrep\ of $\cala_0$.
 
To generalize the relation \erf{45} as well, we associate to any primary field 
$\mu$ a function $\Psi_\mu{:}\ \calg \to \complex$ by
  \be  \Psi_\mu(\J) := \exp(2\pi\ii Q_\J(\mu)) \,.  \ee
{}From the relation $Q_{\J_1}(\J_2\mu)=Q_{\J_1}(\J_2)+Q_{\J_1}(\mu) \bmod\zet $
(which holds because the monodromy charge is additive under operator products)
we learn that $Q_{\J_1}(\mu) +Q_{\J_2}(\mu)=Q_{\J_1\J_2}(\mu) \bmod\zet$,
which in turn implies that the function $\Psi_\mu$ is a group character on
$\calg$. Summing the identity
$\delta_{\J\alpha,\beta}^{} = \sum_\mu S_{\J\alpha,\mu}^{} S^*_{\mu,\beta}
= \sum_\mu \Psi_\mu(\J)\, S_{\alpha,\mu}^{} S^*_{\mu,\beta}$
over $\J\iN\calg$, we therefore obtain
  \be  \Frac1{|\calg|} \sum_{\J\in\calg} \delta_{\J\alpha,\beta}^{}
  = \!\sumnull\mu S_{\alpha,\mu}^{} S^*_{\mu,\beta} \,.  \ee
This settles the generalization from $\zet_2$ to an arbitrary subgroup $\calu$
of $\calg$.
 
\sect{The \fsa}

As the crucial ingredient which allows to obtain a formula for the boundary 
conditions, we now introduce a new $\zet_2$ graded associative algebra 
$\calap$ of dimension $N_0+\Nf$ that contains the \cza\ $\cala_0$ as a 
subalgebra. We claim that this algebra
$\calap$ constitutes the \fsa\ for the case of the modular invariant
\erf{23}. We define $\calap$ as follows.  A distinguished basis of 
$\calap$ is labelled by all possible bulk fields, i.e.\ by the primary fields 
with vanishing monodromy charge and the fixed points. $\cala_0$ is the 
subalgebra of $\calap$ that corresponds to the unit element in the $\zet_2$ grading; the description of the other structure constants $\calnp_{\mu,\nu}
^{\,\ \ \lambda}$ requires some preparation.

Recall that the fusion coefficients $\caln_{\lambda,\mu,\nu}^{}
=\caln_{\lambda,\mu}^{\ \ \nu^+}$ count the (finite) dimension of the spaces of 
chiral blocks of the three-point functions on the sphere. They can be expressed
in terms of the modular matrix $S$ via the Verlinde formula
  \be \caln_{\lambda,\mu,\nu}^{} = \sum_\rho \frac
  {S_{\lambda,\rho} S_{\mu,\rho} S_{\nu,\rho}}{S_{0,\rho}} \, . \ee
The simple current relation \erf{sm} for the entries of $S$ implies that
$\caln_{\lambda,\mu,\nu}^{}\eq\caln_{\J\lambda,\J\mu,\nu}^{}$. 
In fact, the action of a simple current \J\ can be naturally implemented on the 
spaces of chiral blocks, and the latter equality follows from the existence of 
an isomorphism $\tj$ between the respective spaces of blocks. 

Now suppose that both $\lambda\eq f$ and $\mu\eq g$ are fixed points. In this 
case the isomorphism $\tj$ becomes an {\em endo\/}morphism of the space $\cb$
of chiral blocks, and one can compute its trace
  \be  \calno_{f,g,\nu} := {\rm tr}_{\cb}^{}\, \tj \,.  \ee
Note that $\calno_{f,g,\nu}$ is an integer;
in fact, this remains true for the general case of arbitrary simple
current group $\calu$, where the order of \J\ is typically larger than 2. 
In addition, of course, the dimensions of the eigenspaces of $\tj$ are 
non-negative integers; since ${\rm tr}_{\cb}^{}\tj\eq\calno_{f,g,\nu}$
while ${\rm tr}_{\cb}^{}\mbox{\sl id}\eq\caln_{f,g,\nu}$, this means that
  \be  \Frac12\,(\caln_{f,g,\nu} \pm \calno_{f,g,\nu}) \,\in\, \zetpluso \,.
  \labl{cciz}

Similar traces have already appeared in the analysis of the so-called fixed 
point resolution in integer spin simple current modular invariants 
\cite{fusS6}. In fact, it is known \cite{scya6,fusS6}
that there is some other \cft, the so-called fixed point theory, whose primary 
fields are in one-to-one correspondence to the fixed points of the original 
theory (when there is only a single fixed point, as in the \su\ case \erf{inv},
then the fixed point theory is trivial), and whose modular matrix $\chec S$ 
determines $\calno$ via the formula
  \be \calno_{f,g,\nu}^{} = \sumfix h \frac{\chec S_{f,h}\chec S_{g,h}S_{\nu,h}}
  {S_{0,h}}  \,, \labl{53}
where the sum is over all fixed points.\,%
\futnote{It has also been conjectured 
\cite{fusS6} that $\chec S$ describes the modular properties of the one-point 
chiral blocks on the torus, where the insertion is the simple current \J. 
Additional evidence for this relationship has been presented in \cite{bant6}.}
In the case of a \wzwm\ based on an affine \lie\ \g, the fixed point theory
is governed by the so-called orbit Lie algebra $\chec\gM$ that is associated 
\cite{fusS3} to \g\ and \J, which in particular provides an explicit closed 
expression for $\chec S$.

We are now ready to define the multiplication rules for the \fsa\
$\calap$: the product of $Q\eq0$ fields is the ordinary fusion product, while
the other non-vanishing structure constants are given by 
$\calno_{f,g}^{\ \ \lambda}$ and $\calno_{f,\lambda}^{\ \ g}$.\,%
\futnote{As usual, the indices of $\calno$ are raised and lowered by
complex conjugating the corresponding matrix element of $S$ respectively
$\chec S$ on the right hand side of \erf{53}.}
That is,
  \be  \calnp_{\lambda,\mu}^{\,\ \ \nu} = \left\{ \begin{array}{rl}
  \caln_{\lambda,\mu}^{\,\ \ \nu} & {\rm if}\ Q(\lambda)=Q(\mu)=Q(\nu)=0\,,
  \\[.37em]
  \calno_{\lambda,\mu}^{\,\ \ \nu} & {\rm if\; precisely\; one\; out\; of}\
  \lambda,\mu,\nu\ {\rm has}\ Q=0\,, \\[.37em] 0 & {\rm else}\,.
  \eear\right. \ee
Inspection shows that the \fsa\ $\calap$ is commutative and associative, that 
it has $\phi_0$ as a unit element, and that it has a conjugation which is still 
given by the evaluation of the product on $\phi_0$, i.e.
$\calno_{f,g}^{\ \ 0}\eq\delta_{f,g^+}^{}$. As a consequence, $\calap$ 
is again a semisimple associative \alg. However, it is {\em not\/} a fusion 
\alg, because some of its structure constants are negative:
  \be \calno_{f,g,\J\lambda} = \sum_h \Frac{\chec S_{f,h} \chec S_{g,h}
  S_{\J\lambda,h}}{S_{0,h}} = -\, \calno_{f,g,\lambda} \, . \ee
Also note that the algebra $\calap$ is {\em not\/} a subalgebra of the original
fusion algebra $\cala$. 

Applying similar arguments as for $\cala$ above, we can determine the 
$N_0+\Nf= \frac12(N_0+N_1) +2 \Nf$ irreducible representations $\calrp_\alpha$
of $\calap$. They all restrict to irreducible representations of $\cala_0$, 
i.e.\ $\calrp_\alpha(\phi_\mu)\eq S_{\alpha,\mu}/S_{0,\alpha}$ for $Q(\mu)\eq0$.
If $\alpha$ is on a full orbit of \J, then this restriction is uniquely 
extended to the fixed points by zero. On the other hand, by direct calculation 
one checks that for \irrep s corresponding to a fixed point $h$, the extension 
to fixed points is by $\pm\chec S_{h,f}/S_{0,h}$, which accounts for two 
\irrep s $\calrp_{(h+)}$ and $\calrp_{(h-)}$ each. That is,
  \be  \calrp_\alpha(\phi_f^{})=0 \quad{\rm for}\;\ Q(\alpha)\eq0\,, \qquad
  \calrp_{(h\pm)}(\phi_f^{})= \pm\,\Frac{\chec S_{h,f}}{S_{0,h}} \,.  \ee

\sect{The reflection coefficients}

As a consequence of our claim that the \alg\ $\calap$ introduced in the 
previous section governs the boundary 
conditions for a \cft\ with modular invariant \erf{23}, the boundary
coefficients $C^a_{(\mu,\mu^+),0}$ are given by the \irrep\ matrices of 
$\calap$. Hence our results about the \rep\ theory of $\calap$
tell us that there are two different types of boundary conditions: for 
length-two orbits $\alpha$ of \J, we obtain 
  \be  B^\alpha_\mu \equiv C^\alpha_{(\mu,\mu^+),0} 
   = \calrp_\alpha(\phi_\mu) = \left\{\begin{array}{cl}
  \Frac{S_{\alpha,\mu}}{S_{0,\alpha}} & {\rm for}\;\ Q(\mu)\eq0\,, \\{}\\[-.9em]
  0 & {\rm for}\;\ \J\mu\eq\mu   \end{array}\right. \labl4
(because of \erf{sm} this does not depend on 
the choice of representative of the orbit $\alpha$),
while fixed point orbits yield two distinct sets of coefficients:
  \be  B^{(f\pm)}_\mu \equiv C^{(f\pm)}_{(\mu,\mu^+),0}
   = \calrp_{(f\pm)}(\phi_\mu) = \left\{\begin{array}{rl}
  \Frac{S_{f,\mu}}{S_{0,f}} & {\rm for}\;\ Q(\mu)\eq0\,, \\{}\\[-.8em]
  \pm\, \Frac{\chec S_{f,\mu}}{S_{0,f}} & {\rm for}\;\ \J\mu\eq\mu \,.
  \eear\right. \labl5

In the case of \su\ this prescription reproduces the results of \cite{prss3}. 
Also notice that the appearance of the modular matrix $\chec S$ of the fixed 
point theory is rather natural; indeed, fixed points belong to the twisted 
sector, and according to \cite{fusS6} the matrix $\chec S$ governs the modular 
transformations of that sector.

To provide more evidence for our prescription for the boundary coefficients,
we study the annulus amplitude. The latter has the general form \cite{cale}
  \be A_{ab}(t)= \sum_\mu\chii_\mu (\Frac{2\ii}t)\,\llb\Frac{S_{0,\mu}}{S_{0,0}}
  \lrb^{-1} (B^a_\mu\tilde C^a_0)^*_{} B^b_\mu \tilde C^b_0 =
  \sum_\mu A_{ab}^\mu\, \chii_\mu(\Frac{\ii t}2) \,. \labl{61}
Here $a$ and $b$ are the boundary conditions at the two boundaries of the
annulus (i.e.\ each of them can take the values $\alpha$ that label full 
orbits as well as the two values $(f\pm)$ for each fixed point label $f$), 
and $t\iN{\dl R}_{>0}$ is the standard modulus 
of the annulus (the modulus of its covering torus is then $\tau\eq\ii t/2$).
The number $\tilde C^a_0 \equiv \langle\psi_0^{aa}\rangle$ is the 
normalization of the one-point function of the identity on a boundary
of type $a$. The second equality in \erf{61} is obtained by 
a modular transformation and gives the amplitude in the open string channel.

The natural value of $\tilde C^a_0$ that generalizes the expressions for
diagonal \cite{cale} and $D_{\rm odd}$-type \su\ \cite{prss3} theories reads
  \be  C^\alpha_0 = \sqrt2\,S_{0,\alpha}\quad{\rm for}\;\ Q(\alpha)=0\,, \qquad
  C_0^f= S_{0,f}/\sqrt2\quad{\rm for}\;\ \J f=f \,.  \labl c
Inserting the formul\ae\ \erf4, \erf5 and \erf c into the relation \erf{61},
we can determine the tensors $A^\mu_{ab}$; we find 
  \be \bearll
  A^\mu_{\alpha \beta} = 
  \caln^{\ \ \alpha}_{\beta,\mu} + \caln^{\;\ \J\alpha}_{\beta,\mu} \,,\quad &
  A^\mu_{(f\pm)(g\pm)} = \Frac12\,\llb \caln_{f^+,g,\mu}+\calno_{f^+,g,\mu} \lrb
  \,, \\[-.5em]{}\\
  A^\mu_{\alpha(f\pm)} = \caln^{\ \ \alpha}_{f,\mu} \,, &
  A^\mu_{(f\pm)(g\mp)} = \Frac12\,\llb \caln_{f^+,g,\mu}-\calno_{f^+,g,\mu} \lrb
  \,.  \end{array} \labl{A1/2}

We can now present evidence for our prescription \erf4 and \erf5. We first 
remark that the annulus amplitude can be regarded as the partition function for
the boundary operators (before orientifold projection). For consistency it is 
therefore necessary that all coefficients $A^\mu_{ab}$ in \erf{A1/2} are 
non-negative integers. Inspection shows that this highly non-trivial constraint
is indeed satisfied for all values of $a,b$ and $\mu$; in the particular case 
of boundary conditions of fixed point type this is a consequence of the result 
\erf{cciz}, which in turn has its origin in the specific properties
of the fixed point theories.

Further confirmation is provided by the following properties of the tensors 
$A^\mu_{ab}$, which we deduce from the formul\ae\ \erf{A1/2}.
First, they obey the relation
  \be  \sum_\mu A^\mu_{ab} A^{\mu^+}_{cd}
  = \sum_\mu A^\mu_{ac^+} A^{\mu^+}_{b^+d}  \labl{4a}
for all choices of the boundary labels $a,b,c,d$ 
(also note that $A^\mu_{ba}\eq A^\mu_{a^+b^+}$). And second,
considered as matrices in the boundary labels, they satisfy
  \be  A^\mu A^\nu = \sum_\lambda \caln_{\mu,\nu}^{\ \ \lambda} A^\lambda \,.
  \labl{aaa}
(For diagonal theories, where $A^\mu_{\alpha\beta}\eq\caln^{\ \ \alpha}_{\beta,
\mu}$, these relations reduce to the statement that the fusion rules are
associative and that the structure constants furnish a \rep\
-- the regular representation -- of the fusion \alg.) 
The equality \erf{aaa} can be interpreted as the assertion that the boundary 
conditions are complete. More specifically (compare equation (33) of 
\cite{prss3}), it implies that the two distinct ways of factorizing a two-point
function with bulk insertions lead to the same result.

It is known \cite{prss3} that the relations \erf{4a} and \erf{aaa} are highly
restrictive, in particular when they are combined with the information 
about the spectrum that is contained in the torus partition function. The fact 
that our ansatz for the boundary coefficients reproduces these formul\ae\ 
is therefore another strong indication that our prescription is correct.

As a final test, we study the boundaries along the lines followed
in \cite{card9} for diagonal modular invariants. We first remark 
that when both boundaries of the annulus are in the $\alpha\eq0$ condition, 
then, in the terminology of \cite{card9}, both the field $\phi_0$ and $\phi_\J$ 
propagate in the bulk. This nicely fits with the observation of \cite{prss3} 
that there is an effective enhancement of the boundary symmetry. Now just like 
in the case of integer spin simple current invariants, for fixed points such 
an extension can be performed in two inequivalent ways. Therefore for boundary 
conditions $a=0$ and $b=(f\pm)$ $\phi_f$ propagates in the bulk, and it does so
in two independent ways (as characters of the {\em non-extended\/} algebra, the
associated characters are, however, identical, 
$\chii_{(f+)}=\chii_{(f-)}=\chii_f$).
 
We now map the annulus to an infinitely long strip and consider the following
configuration (compare figure 2 of \cite{card9}). We start with conditions of 
type $\alpha=0$ both on the left and on the right boundary of the strip; then 
we insert on the right boundary 
a boundary operator that switches to boundary conditions $(f\pm)$ 
so that now $\phi_f$ propagates in the bulk. Afterwards we insert a boundary 
operator on the left boundary that yields boundary condition $(g\pm)$. This 
amounts to coupling the two fixed point primary fields, and we know that they 
can only couple to primary fields with vanishing monodromy charge.
The latter are, however, `uncharged' under the action of the simple current,
and accordingly we get a restriction from the requirement that the couplings 
transform correctly under the simple current action. More specifically, if the 
couplings $(f\pm)$ and $(g\pm)$ are of like sign, then the coupling should be 
even; as we have seen in the discussion before equation \erf{cciz}, the number 
of such couplings is just $\frac12(\caln_{f,g}^{\ \ \mu}\pl\calno_{f,g}^{\ \ 
\mu})$.  Similarly, the case of opposite signs yields 
$\frac12(\caln_{f,g}^{\ \ \mu} \mi\calno_{f,g}^{\ \ \mu})$ couplings. Our 
argument thus reproduces precisely what we obtained in the second column of 
\erf{A1/2}. It would be gratifying to corroborate this generalization of the 
rather heuristic arguments of \cite{card9} by an explicit calculation analogous 
to the one reported in \cite{prss3}. The latter, however, relies on the explicit
knowledge of the duality matrices, which are available for \su, but not for more
complicated \cfts. 

\sect{Conclusions}

In this letter, we have determined the boundary conditions for \cfts\ with 
non-trivial modular invariants of `$D_{\rm odd}$-type' \erf{23}.
We have shown that just as in the diagonal case they are
controlled by a semisimple algebra, the \fsa\ $\calap$.
The structure we discovered is closely related to the fusion \alg\ of
another type of modular invariants, namely those of `$D_{\rm even}$-type' 
(also known as integer spin simple current extensions).
In particular, it looks as if the boundary
theory is extended by the {\em half\/}-integer spin simple current \J. 

This is indeed most remarkable, because in the case of extensions, modular 
invariance provides powerful consistency requirements.
But for surfaces of Euler characteristic zero which are non-orientable
or which have a boundary, there is no analogue of a modular group. In string 
theory it is usually argued that tadpole cancellation provides a substitute 
for such consistency conditions. Note, however, that for the investigations 
presented in this letter we did not have to assume that the \cft\ is part of
a string compactification (e.g., the central charge is not restricted), so that 
the conditions of tadpole cancellation cannot even be formulated.
Still, it seems that already on a pure \cft\ level there are similar powerful 
constraints; to unravel the underlying structure will be a promising task.

As a final and somewhat more speculative remark, we mention that the quantum 
dimensions, and hence also the space of possible boundary conditions, carry
a natural action of the Galois group of a cyclotomic number field.
When a \cft\ admits a geometrical interpretation as a sigma model 
on some manifold $\cal M$, then a boundary condition frequently 
corresponds to a certain sheaf on $\cal M$. If those sheaves are the direct 
image of a line bundle over a spectral cover, one might speculate on a relation
between the Galois group of the corresponding covering and the Galois action 
just mentioned.

\bigskip
\bigskip\noindent
{\bf Acknowledgement:}\\ We are grateful to A.\ Sagnotti, A.N.\ Schellekens
and Ya.S.\ Stanev for helpful discussions.
\vfill

 \newcommand\wb{\,\linebreak[0]} \def\wB {$\,$\wb}
 \newcommand\Bi[1]    {\bibitem{#1}}
 \newcommand\Erra[3]  {\,[{\em ibid.}\ {#1} ({#2}) {#3}, {\em Erratum}]}
 \newcommand\BOOK[4]  {{\em #1\/} ({#2}, {#3} {#4})}
 \renewcommand\J[5]   {\ {\sl #5}, {#1} {#2} ({#3}) {#4} }
 \newcommand\Prep[2]  {{\sl #2}, preprint {#1}}
 \def\comp  {Com\-mun.\wb Math.\wb Phys.}
 \def\ijmp  {Int.\wb J.\wb Mod.\wb Phys.\ A}
 \def\mpla  {Mod.\wb Phys.\wb Lett.\ A}
 \def\nupb  {Nucl.\wb Phys.\ B}
 \def\phlb  {Phys.\wb Lett.\ B}
\def\furu    {fusion rule}
\def\modinv  {modular invarian}

\end{document}